\documentstyle[11pt]{article}
\begin{document}

%  Greek letters

\def\a{\alpha}
\def\b{\beta}
\def\d{\delta}
\def\e{\epsilon}
\def\g{\gamma}
\def\k{\kappa}
\def\l{\lambda}
\def\o{\omega}
\def\t{\theta}
\def\s{\sigma}
\def\z{{\bar{z}}}
\def\D{\Delta}
\def\L{\Lambda}
\def\T{\Theta}

\def\p{\partial}
\def\cp2{\cos\frac{\phi}{2}}

\def\cL{{\cal L}}
\def\cB{{\cal B}}

% Shorthands for \begin{equation} and the like

\def\beq{\begin{equation}}
\def\eeq{\end{equation}}
\def\bea{\begin{eqnarray}}
\def\eea{\end{eqnarray}}
\def\ba{\begin{array}}
\def\ea{\end{array}}
\def\no{\nonumber}
\def\lt{\left}
\def\rt{\right}
\newcommand{\bq}{\begin{quote}}
\newcommand{\eq}{\end{quote}}

\begin{titlepage}
\begin{flushright}
YITP-95-16\\
hep-th/9512085
\end{flushright}
\vskip.3in
\begin{center}
{\huge Integrable Four-Fermi Models with a Boundary and 
Boson-Fermion Duality}
\vskip.3in
{\Large Takeo Inami, Hitoshi Konno} 
and 
{\Large Yao-Zhong Zhang~\footnote{ Address after June 6, 1996:
Department of Mathematics, University of Queensland, Brisbane, Australia.
Email: yzzhang@yukawa.kyoto-u.ac.jp; yzz@maths.uq.oz.au.}}
\vskip.1in
{\em Yukawa Institute for Theoretical Physics\\
Kyoto University, Kyoto 606, Japan}

\end{center}
\vskip.4in
\begin{center}
{\bf Abstract}
\end{center}
\begin{quote}

Construction of integrable field theories in space with a boundary
is extended to fermionic
models. We obtain general forms of boundary interactions consistent with
integrability of the massive Thirring model and study the duality 
equivalence of the MT model and the sine-Gordon model with boundary
terms. We find a variety of
integrable boundary interactions in the $O(3)$ Gross-Neveu model from the
boundary supersymmetric sine-Gordon theory by using boson-fermion duality.

\end{quote}
%\vskip 1.2cm
%\noindent {\bf PACS numbers:} 02.20.+b, 02.40.+m, 05.50.+q

\end{titlepage}
\newpage

\newcommand{\sect}[1]{\setcounter{equation}{0}\section{#1}}
\renewcommand{\theequation}{\thesection.\arabic{equation}}

\sect{Introduction\label{intro}}

A physical system can often be described in two apparently different languages,
which are dual to each other. One of the examples, which has recently 
attracted much attention, is the conjecture of electric-magnetic duality in
supersymmetric Yang-Mills theories in four dimensions \cite{Mon77}. 
In two dimensional
quantum field theories (QFTs), the duality relation between the bosonic
and fermionic descriptions can be proven in a large class of integrable
models, e.g. the  sine-Gordon (SG) and massive Thirring (MT) equivalence
\cite{Col75} and the WZNW and massless Thirring equivalence \cite{Wit84}. 
Studies of the duality relation have in the past deepened
our understanding of the non-perturbative nature of QFTs.

Integrable QFTs including SG and supersymmetric sine-Gordon (SSG) theories have
recently been extended to cases in which the space has a boundary
\cite{Che84,Car89,Gho94,Cor94,Sal95,Ina95,Ahn95}. Integrable boundary QFT 
can only be constructed if one can
find boundary interactions which satisfy suitable conditions. Integrable
boundary QFTs have many applications in physical systems of low dimensionality.

A large class of integrable QFTs consisting of fermions is known.
It is believed that such models bear a resemblance, in some respects such
as chiral symmetry breaking, to the
fermions in gauge theories in four dimensions.

In this paper we extend the construction of integrable boundary QFT to
fermionic models and
study boson-fermion duality in models with a boundary potential.
We consider the two simplest cases,  
MT and $O(3)$ Gross-Neveu (GN) models \cite{Zam79}.
In the bulk theory, the MT and the $O(3)$ GN models are known to be 
bosonized to give the SG and the SSG theories \cite{Col75,Wit78}, respectively.
We will find a few different classes
of integrable boundary interactions, given a 
fermionic model in the bulk, each of them defining a distinct boundary QFT.

\sect{Four-Fermi Interaction  Models and Their Dual Theories}

Let us recall in this section some known facts about the
four-fermi interaction models in the bulk and their duals.

The MT model is defined by \footnote{Throughout this paper, we use the
following notation: $\bar{\psi}=\psi^\dagger\g^0,~\g^0=\lt (
\begin{array}{cc}
0 & 1\\
-1 &0\\
\end{array}   \rt),~~
\g^1=\lt(
\begin{array}{cc}
0 & 1\\
1 &0\\
\end{array}   
\rt)$ and $\g_\mu=\eta_{\mu\nu}\g^\nu,~\eta_{\mu\nu}=diag(-1,1)$.}
\beq
\cL_{0MT}=\frac{1}{2}\bar{\psi}i\g^\mu\p_\mu\psi-\frac{1}{2}\p_\mu\bar{\psi}
   i\g^\mu\psi-im\,\bar{\psi}\psi-\frac{g}{2}
   \lt(\bar{\psi}\g_\mu\psi\rt)^2,\label{l0-mth}
\eeq
where $\psi$ is a complex spinor (Dirac fermion), $m$ a mass
parameter and $g$ a coupling constant.

The MT model is  known to be equivalent to
the sine-Gordon theory \cite{Col75}, which is a model of a single
real scalar field $\phi(x)$ defined by
\beq
\cL_{0SG}=\frac{1}{2}\left(\p_\mu\phi\right)^2
  +\frac{m^2_0}{\s^2}\cos\s\phi.\label{l0-sg}
\eeq
The coupling constants in the two theories are connected by
\beq
g/\pi=4\pi/\s^2-1.
\eeq
This equivalence between the SG theory and the MT model has been shown to
be a duality relation \cite{Bur94}.

The $O(3)$ GN  model is a model of a triplet of  real
spinors (Majorana fermions) $\chi_i(x)$ and is
defined by 
\beq
{\cal L}_{0GN}=-\frac{1}{2}\sum_{i=1}^3\bar{\chi}_i
   i\g^\mu\p_\mu\chi_i
   -g\lt(\sum_{i=1}^3\bar{\chi}_i\chi_i\rt)^2.\label{l0-gn}
\eeq
This model was shown by Witten \cite{Wit78} to be equivalent to the
SSG model which is defined by the lagrangian
\beq
{\cal L}_{0SSG}=-\frac{1}{2}\left(\p_\mu\phi\right)^2
  -\frac{1}{2}\bar{\chi}i\g^\mu\p_\mu\chi
  -2A^2\cos^2B\phi-iAB\,\bar{\chi}\chi\cos B\phi,\label{s0-ssg}
\eeq
where $\chi$ is a Majorana spinor 
and $A,~B$ are coupling constants.
The point is to group the Majorana fermions in pairs. Given two Majorana
spinors $\chi_1$ and $\chi_2$ (two Majorana fermions being equivalent to
a Dirac fermion), one has the following fermion-boson 
correspondence \cite{Wit78}
\bea
&&\frac{1}{2}(\bar{\chi}_1i\g^\mu\p_\mu\chi_1+\bar{\chi}_2i\g^\mu\p_\mu
   \chi_2)=\frac{1}{2}(\p_\mu\phi)^2,\no\\
&&i\bar{\chi}_1\chi_1+i\bar{\chi}_2\chi_2=:\cos\sqrt{4\pi}\phi:,\no\\
&&\bar{\chi}_1\g^\mu\chi_2=-\frac{1}{\sqrt{\pi}}\e^{\mu\nu}
   \p_\nu\phi,\label{bf-relation1}
\eea
together with the following important relation 
\beq
:\cos\sqrt{4\pi}\phi:^2=\frac{-2}{\pi}(\p_\mu\phi)^2+ 
  c~{\rm number},\label{bf-relation2}
\eeq
where :: stands for normal ordering. The last
relation can be deduced from (\ref{bf-relation1}) and 
has no classical counterpart.

Using these fermion-boson relations, one gets \cite{Wit78}
\beq
{\cal L}_{0GN}=-\frac{1}{2}(\p_\mu\varphi)^2-
           \frac{1}{2}\bar{\chi}_3i\g^\mu\p_\mu
           \chi_3-2A^2:\cos\rho\varphi:^2-iA\rho\,\bar{\chi}_3
           \chi_3:\cos\rho\varphi:,\label{l0-gn-bosonized}
\eeq
where
\bea
&&A=\sqrt{C/2},~~~~\rho=2g/A,~~~~\varphi=\sqrt{\pi}\frac{A}{g}\phi,\no\\
&&C=2g^2(1-4g/\pi)/(\pi-8g^2/\pi).\label{funny-constants}
\eea
The coupling 
constant $A$ is connected to the coupling constant $g$ by the above relations 
(\ref{funny-constants}) and $B$  to $g$ by
\beq
B=2g/A.
\eeq
The equivalence of the two descriptions (\ref{l0-gn}) and 
(\ref{l0-gn-bosonized}) may be interpreted as a duality relation, in the same
way as the MT--SG equivalence.

\sect{Integrable MT Model on a Half-Line\label{mt-integrability}}

We use the light-cone notation: 
$z=\frac{1}{2}(t-x),~\bar{z}=\frac{1}{2}(t+x)$, which implies
$\p_z=\p_t-\p_x,~\p_{\bar{z}}=\p_t+\p_x$. The Dirac fermion $\psi$ appearing
in (\ref{l0-mth}) can be expressed as $\psi=(\psi_1,\psi_2)^T$.

The theory on a half-line $x\in (-\infty, 0]$ is defined by adding the boundary
term $S_b$ to the bulk part $S_0$ of the action.
The total action $S$ can be written as
\beq
S=S_0+S_b\equiv \int_{-\infty}^\infty dt\int_{-\infty}^0dx\;{\cal L}_0
  (\psi_1,\psi_2,\p_z\psi_1,\p_{\bar{z}}\psi_2)
  +\int_{-\infty}^\infty dt\; {\cal B}(\psi_1, \psi_2),\label{gen-s}
\eeq
where the boundary potential
$\cB$ is assumed to be a functional of the fields at $x=0$ but not 
of their derivatives.
In addition to the bulk field equations:
$\frac{\d{\cal L}_0}{\d\psi_1}-\p_z\frac{\d{\cal L}_0}
   {\d\p_z\psi_1}=0,~~
\frac{\d{\cal L}_0}{\d\psi_2}-\p_{\bar{z}}\frac{\d{\cal L}_0}
   {\d\p_{\bar{z}}\psi_1}=0$,
we have equations of motion at the boundary $x=0:~
-\frac{\d{\cal L}_0}{\d\p_z\psi_1}+\frac{\p\cB}{\p\psi_1}=0,~~
\frac{\d{\cal L}_0}{\d\p_{\bar{z}}\psi_2}+\frac{\p\cB}{\p\psi_2}=0$.
 
In the case of MT model, if one makes the substitution (rescaling):
\beq
\frac{1}{m}\p_\pm\longrightarrow\p_\pm,~~~~\sqrt{\frac{2g}{m}}\psi_1
   \longrightarrow\psi_1,~~~~\sqrt{\frac{2g}{m}}\psi_2\longrightarrow  i\psi_2,
\eeq
then from (\ref{l0-mth}) one has the following lagrangian for the $\cL_0$'s
in (\ref{gen-s}):
\beq
{\cal L}_{0MT}=-\frac{i}{2}\psi_1^\dagger\p_z\psi_1+\frac{i}{2}\p_z
      \psi_1^\dagger\psi_1-\frac{i}{2}\psi_2^\dagger\p_{\bar{z}}\psi_2
      +\frac{i}{2}\p_{\bar{z}}\psi_2^\dagger\psi_2
      +\psi_1^\dagger\psi_2+\psi_2^\dagger\psi_1
      +\psi_1^\dagger\psi_1\psi_2^\dagger\psi_2.\label{L-thm}
\eeq
The equations of motion read in the bulk
\bea
i\p_z\psi_1&=&\psi_2+\psi_1\psi_2^\dagger\psi_2,\no\\
i\p_{\bar{z}}\psi_2&=&\psi_1+\psi_2\psi_1^\dagger\psi_1,\label{eqm-mth}
\eea
and at the boundary $x=0$:
\beq
-i\psi_1+\frac{\p\cB}{\p\psi_1^\dagger}=0,~~~~~~
i\psi_2+\frac{\p\cB}{\p\psi_2^\dagger}=0.\label{bc-mth}
\eeq

In the bulk theory there is an infinite number of conserved
charges \cite{Ber76}
constructed from densities $T_{s+1},~\bar{T}_{s+1},~
\T_{s-1}$ and $\bar{\T}_{s-1}$ with $s=1,3,5,\cdots$.
These densities satisfy the following continuity equations 
\beq
\p_\z T_{s+1}=\p_z\T_{s-1}\,,~~~~~~\p_z\bar{T}_{s+1}=\p_\z\bar{\T}_{s-1}.
\eeq
The densities for $s=1$  are given by the energy-momentum tensor,
\bea
&&\bar{T}_2=i\psi_1^\dagger\p_{\bar{z}}\psi_1-i\p_{\bar{z}}
   \psi_1^\dagger\psi_1,\no\\
&&T_2=i\psi_2^\dagger\p_z\psi_2-i\p_z\psi_2^\dagger\psi_2,\no\\
&&\T_0=\bar{\T}_0=-\psi_2^\dagger\psi_1-\psi_1^\dagger\psi_2.
\eea
The $s=3$ densities are 
\bea
\bar{T}_4&=&-i\psi_1^\dagger\p_{\bar{z}}^3\psi_1+
     i\p_{\bar{z}}^3\psi_1^\dagger\psi_1+
     6\p_{\bar{z}}\psi_1^\dagger\psi_1^\dagger\psi_1\p_{\bar{z}}\psi_1,\no\\
T_4&=&-i\psi_2^\dagger\p_z^3\psi_2+i\p_z^3
     \psi_2^\dagger\psi_1+6
     \p_z\psi_2^\dagger\psi_2^\dagger\psi_2\p_z\psi_2,\no\\
\bar{\T}_2&=&\psi_2^\dagger\p_{\bar{z}}^2\psi_1+\p_{\bar{z}}^2\psi_1^\dagger
      \psi_2-i\psi_2^\dagger\psi_1^\dagger\psi_1\p_{\bar{z}}\psi_1
      +i\p_{\bar{z}}\psi_1^\dagger\psi_1^\dagger\psi_1\psi_2,\no\\
\T_2&=&\psi_1^\dagger\p_z^2\psi_2+\p_z^2
      \psi_2^\dagger\psi_1
      -i\psi_1^\dagger\psi_2^\dagger\psi_2\p_z\psi_2
      +i\p_z\psi_2^\dagger\psi_2^\dagger\psi_2\psi_1
     .\label{theta-thetabar}
\eea
These densities can be checked to satisfy the continuity equations, using only
the field equations (\ref{eqm-mth})
 and the anticommutation relations for the $\psi$'s and
their derivatives.

Suppose that one can choose the boundary potential $\cB$ such that at $x=0$ 
\beq
T_4-\bar{T}_4+\T_2-\bar{\T}_2=\frac{d}{dt}\Sigma_3(t),\label{sigma}
\eeq
where $\Sigma_3(t)$ is some functional of boundary 
fields $\psi_1(t),~\psi_2(t)$.
Then the charge $P_3$, given by
\beq
P_3=\int^0_{-\infty}dx\;(
T_4+\bar{T}_4-\T_2-\bar{\T}_2)-\Sigma_3(t)
\eeq
is a non-trivial integral of motion.

We now examine in what circumstances  $T_4-\bar{T}_4+\T_2-\bar{\T}_2$ may
be written as a total $t$-derivative.
By using the field equations (\ref{eqm-mth}) and the anticommutation
property of the $\psi$'s, it can be shown after a tedious
computation that
%\bea
%\psi_{1x}&=&i(\psi_2+\psi_1\psi_2^\dagger\psi_2)+\psi_{1t},\no\\
%\psi_{1xx}&=&\psi_1-i\psi_{2t}+2\psi_2\psi_1^\dagger\psi_1
%   +i\psi_{1t}\psi_2^\dagger\psi_2
%   -i\psi_1\psi_{2t}^\dagger\psi_2\no\\
%& &-i\psi_1\psi_2^\dagger\psi_{2t}
%   +i\p_t(\psi_2+\psi_1\psi_2^\dagger\psi_2)+\psi_{1tt},\no\\
%\psi_{1xxx}&=&i\psi_2+i\psi_{2tt}+3i\psi_1\psi_2^\dagger\psi_2
%   -4\psi_{2t}\psi_1^\dagger\psi_1-2\psi_{1t}\psi_1^\dagger\psi_2
%   -2\psi_{2t}\psi_2^\dagger\psi_2\no\\
%& &+\psi_{1t}\psi_2^\dagger\psi_1
%   -\psi_1\psi_2^\dagger\psi_{1t}
%  +i\psi_{1tt}\psi_2^\dagger\psi_2
%   -2i\psi_{1t}\psi_{2t}^\dagger\psi_2\no\\
%& &-2i\psi_{1t}\psi_2^\dagger\psi_{2t}+i\psi_1\psi_{2tt}^\dagger\psi_2
%   +2i\psi_1\psi_{2t}^\dagger\psi_{2t}+i\psi_1\psi_2^\dagger\psi_{2tt}\no\\
%& &+\p_t(\psi_1-i\psi_{2t}+2\psi_2\psi_1^\dagger\psi_1+i\psi_{1t}\psi_2^\dagger
%    \psi_2-i\psi_1\psi_{2t}^\dagger\psi_2-i\psi_1\psi_2^\dagger\psi_{2t})\no\\
%& &+i\p_t^2(\psi_2+\psi_1\psi_2^\dagger\psi_2)+\psi_{1ttt},\label{xxx}
%\eea
%where  $Q_x,~Q_{xt}$ etc stand for $\p_xQ,~\p_x\p_tQ$ and so on.
%Similar relations for $\psi_{2x},~\psi_{2xx}$ and $\psi_{2xxx}$ can be
%obtained.
% 
%Using (\ref{xxx}) and the similar expressions for $\psi_{2x},~\psi_{2xx}$
%and $\psi_{2xxx}$, and after a long but direct computation, one gets
\bea
T_4-\bar{T_4}+\T_2-\bar{\T}_2&=&\p_t({\rm something})-
       24(\psi_{1t}^\dagger\psi_{1t}+\psi_{2t}^\dagger\psi_{2t})
       (\psi_1^\dagger\psi_1-\psi_2^\dagger\psi_2)\no\\
& & -24i\,\psi_1^\dagger\psi_1(\psi_{1t}^\dagger\psi_2-\psi_2^\dagger
    \psi_{1t})-24i\,\psi_2^\dagger\psi_2(\psi_1^\dagger\psi_{2t}
    -\psi_{2t}^\dagger\psi_1)\no\\
& &-16i\,(\psi_{1t}^\dagger\psi_{1tt}-\psi_{2t}^\dagger\psi_{2tt})
    -12i\,(\psi_{1t}^\dagger\psi_1-\psi_{2t}^\dagger\psi_2).\label{t4-t4bar}
\eea

We look for the boundary potential ${\cal B}(\psi_1,\psi_2)$ which will make
the r.h.s of (\ref{t4-t4bar}) be a total $t$-derivative.
The most general form of $\cB(\psi_1,\psi_2)$ satisfying hermiticity
must have the form \footnote{We have excluded terms such as 
$\psi_1^\dagger\psi_1\psi_2^\dagger\psi_2,~\psi_1^\dagger\psi_1\psi_2^\dagger$
etc, since they do not have the right dimensions.}
\bea
{\cal B}(\psi_1,\psi_2)&=&a_1\psi_1^\dagger\psi_1+
         a_2\psi_2^\dagger\psi_2+ia_3\lt(\psi_1^\dagger\psi_2^\dagger-
         \psi_2\psi_1\rt)+ia_4\lt(e^{ib_1}
           \psi_1^\dagger\psi_2-e^{-ib_1}\psi_2^\dagger\psi_1\rt)\no\\
& & +a_5\lt(\e_1^\dagger\psi_1+\psi_1^\dagger\e_1\rt) 
          +a_6\lt(e^{ib_2}\e_2^\dagger\psi_2+e^{-ib_2}\psi_2^\dagger
          \e_2\rt)\no\\
& & +ia_7\e_1^\dagger\p_t\e_1+ia_8\e_2^\dagger\p_t\e_2+ia_9\e_1^\dagger
     \p_t\e_2,\label{bc-ansatz}
\eea
where $a_l~(l=1,\cdots,9),~b_1,~b_2$ are real constant bosonic parameters,
and $\e_1(t),~\e_2(t)$ are fermionic boundary operators \cite{Gho94}.
Then, the $\psi_1,~\psi_2$ field equations at the boundary
$x=0$, eqs.(\ref{bc-mth}),  become
\bea
&&(1+ia_1)\psi_1-a_4e^{ib_1}\psi_2-a_3\psi_2^\dagger=
   -ia_5\e_1,\no\\
&&(1-ia_2)\psi_2-a_4e^{-ib_1}\psi_1-a_3\psi_1^\dagger
   =ia_6e^{-ib_2}\e_2.\label{bc-eqm-field}
\eea

It turns out that in order for solutions of (\ref{bc-eqm-field})
to make the r.h.s of (\ref{t4-t4bar})
a total $t$-derivative one has to set $\e_1(t)=\e_2(t)\equiv\e(t)$.
The constraints for the other constant parameters
can also be determined.
We find the following four classes of boundary
potential which preserve the integrability of the bulk MT model:
\bea
(i)~~{\cal B}(\psi_1,\psi_2)&=&\frac{\sin\a_0}{\cos\a_0}\lt(\psi_1^\dagger\psi_1
         -\psi_2^\dagger\psi_2\rt)+\frac{i}{\cos\a_0}
         \lt(\psi_1^\dagger\psi_2^\dagger
         -\psi_2\psi_1\rt),\no\\
& &      \a_0\neq \frac{\pi}{2}~{\rm mod~} \pi;\label{solution1}\\
(ii)~~{\cal B}(\psi_1,\psi_2)&=&\frac{\sin(\b-\b_0)}{\cos(\b-\b_0)}
         \lt(\psi_1^\dagger\psi_1+\psi_2^\dagger\psi_2\rt)\no\\
& &+\frac{i}{\cos(\b-\b_0)}\lt(e^{i\b}\psi_1^\dagger\psi_2
         -e^{-i\b}\psi_2^\dagger\psi_1\rt),\no\\
& &      \b-\b_0\neq\frac{\pi}{2}~{\rm mod}~\pi;\label{solution2}\\
(iii)~~{\cal B}(\psi_1,\psi_2)&=&a(\psi_1^\dagger\psi_1-
           \psi_2^\dagger\psi_2)+if\lt (
           \psi_1^\dagger\psi_2^\dagger-\psi_2\psi_1\rt)\no\\
& & +h\lt(\e^\dagger\psi_1+ \psi_1^\dagger\e
    +e^{i\g}\e^\dagger\psi_2+e^{-i\g}\psi_2^\dagger\e\rt )
    +ib\e^\dagger\p_t\e,\no\\
& &   f\neq\pm 1,~{\rm if~} a=0,{\rm ~and~}
   a\neq 0~{\rm if}~f=\pm 1 ;\label{solution3}\\
(iv)~~{\cal B}(\psi_1,\psi_2)&=&r(\psi_1^\dagger\psi_1+
           \psi_2^\dagger\psi_2)+ic\lt (e^{i\a}
           \psi_1^\dagger\psi_2-e^{-i\a}\psi_2^\dagger\psi_1\rt )\no\\
& & +s\lt(\e^\dagger\psi_1+ \psi_1^\dagger\e
    +e^{i\l}\e^\dagger\psi_2+e^{-i\l}\psi_2^\dagger\e\rt )
    +id\e^\dagger\p_t\e,\no\\
& &   c\neq\pm 1,~{\rm if~} r=0,{\rm ~and~}
   r\neq 0~{\rm if}~c=\pm 1,\label{solution4}
\eea
where $\a_0,~\b_0,~\b_,~\a,,~\g,~\l$, and $a,~b$ etc, are free 
real constant parameters.
It is remarkable that in the cases $(iii)$ and $(iv)$, five and six
free parameters are allowed 
in the integrable boundary potential, respectively. 

These boundary potentials have been derived
by examining the first non-trivial conserved
charge $P_3$ and our argument is restricted to the classical case. 
We believe that the above
computation can be extended to the quantum theory, and
that our analysis will be completed by showing
that all conserved charges of higher spin give the same results. To this
end one may use the prescription suggested in \cite{Bow95}.

The boundary conditions for the $\psi$ fields can be derived from the above 
boundary potentials. The boundary term (\ref{solution1}) gives rise to the
free boundary condition, 
\beq
(i)~~~\psi_1=e^{i\a_0}\psi_2^\dagger,\label{free-bc}
\eeq
%\footnote{In \cite{LeC95}, the authors introduced
%an unnecessary phase in the free boundary condition, which can
%be absorbed into the definition of the fields $\psi_1$ and $\psi_2$.}
and (\ref{solution2}) to the fixed boundary condition,
\beq
(ii)~~~\psi_1=e^{i\b_0}\psi_2.\label{fixed-bc}
\eeq
In the last two cases one needs to eliminate
the ``external field" $\e(t)$  by using the $\e$ field equation at the boundary
$x=0$, which takes the form for the case $(iii)$,
$ib\p_t\e+h(\psi_1+e^{i\g}\psi_2)=0$,
and $id\p_t\e+s(\psi_1+e^{i\l}\psi_2)=0$
for the case $(iv)$. The results are
\bea
(iii)~~~f\psi_2^\dagger-(1+ia)\psi_1-e^{i\g}\lt( (1+ia)\psi_2
        -f\psi_1^\dagger\rt)&=&0,\no\\
f\p_t\lt(\psi_2^\dagger-(1+ia)\psi_1\rt)+h^2\lt(\psi_1+e^{i\g}\psi_2\rt)
     &=&0;\label{bc-iii}\\
(iv)~~~ce^{i\a}\psi_2-(1+ir)\psi_1-e^{i\l}\lt((1-ir)\psi_2-ce^{-i\a}
         \psi_1\rt)&=&0,\no\\
d\p_t\lt(ce^{i\a}\psi_2-(1+ir)\psi_1\rt)+s^2\lt(\psi_1+e^{i\l}\psi_2\rt)
      &=&0,\label{bc-iv}
\eea
plus the corresponding equations obtained by hermitian conjugation.

Let us examine whether the integrable boundary
MT models obtained above have a dual description in terms of bosons. 
Consider the special case of $(iii)$ (the argument below applies to
the case $(iv)$ as well),
\beq
{\cal B}(\psi_1,\psi_2)=
  h\lt(\e^\dagger\psi_1+ \psi_1^\dagger\e
    +e^{i\g}\e^\dagger\psi_2+e^{-i\g}\psi_2^\dagger\e\rt )
\eeq
which can be rewritten as, after a redefinition of the field $\e$:
$\e=\varepsilon e^{-i\d}$,
\beq
{\cal B}(\psi_1,\psi_2)=
  h\lt(e^{i\d}\varepsilon^\dagger\psi_1+ e^{-i\d}\psi_1^\dagger\varepsilon
    +e^{i(\g+\d)}\varepsilon^\dagger\psi_2+
   e^{-i(\g+\d)}\psi_2^\dagger\varepsilon\rt ).\label{s-solution}
\eeq
With the help of a result obtained in \cite{LeC95}, one may write
\beq
:\cos\frac{\s(\phi-\phi_0)}{2}:
   =e^{i\s\phi_0/2}(\varepsilon^\dagger\psi_1+\varepsilon\psi_2^\dagger)
    +e^{-i\s\phi_0/2}(\psi_1^\dagger\varepsilon+
    \psi_2\varepsilon^\dagger).\label{boson-fermion}
\eeq
So, the boundary MT model with (\ref{s-solution}) as its boundary potential
is equivalent to the boundary
SG theory \cite{Gho94} which is defined by the lagrangian,
\beq
\cL_{0SG}=\frac{1}{2}\left(\p_\mu\phi\right)^2
  +\frac{m^2}{\s^2}:\cos\s\phi:+\d(x)\,\L\;:\cos\frac{\s(\phi-\phi_0)}{2}:
  ,\label{b-sg}
\eeq
provided the coupling constants
in the two theories are connected by
\beq
h=\L,~~~\d=\frac{\s\phi_0}{2},~~~\g=\pi-\s\phi_0.
\eeq

Other types of boundary potential in the MT model do not appear to give rise 
to integrable boundary potentials in the SG model by means of bosonization.
The reason why bosonization formula, which holds in the bulk theory, may not
hold in some cases with a boundary is not very clear to us. The following
may be a possible explanation. The bosonization involves an exponential of
integral from $-\infty~(+\infty)$ to $x$. The presence of a boundary (at
$x=0$) may be an obstruction to this non-local expression and may cause the
breakdown of boson-fermion duality in boundary integrable field theories.

\sect{Integrable $O(3)$ GN Model on a Half-Line via 
   Boson-Fermion Duality\label{gn-integrability}}

The boundary potential compatible with integrability of the SSG
has been investigated previously \cite{Ina95}. Two different forms of
potentials are obtained:
\bea
&& 1)~~~\cB(\phi,\chi,\bar{\chi})=\L:\cos B(\phi-\phi_0):
        +\frac{i}{4}M\,\bar{\chi}\chi,~~(M\neq\pm 1)
        ; \label{bc-ssg-nonsusy}\\
&& 2)~~~\cB(\phi,\chi,\bar{\chi})=\pm \lt(\frac{2A}{B}:\cos B\phi:
        + \frac{i}{4}\bar{\chi}\chi\rt),\label{bc-ssg-susy}
\eea
where $\L,~M$ and $\phi_0$ are bosonic constants,
and $A,~B$ are the same constants as in 
(\ref{s0-ssg}). Only the boundary potential of the form $2)$ preserves
supersymmetry.

In this section we determine the integrable boundary $O(3)$ GN
model which is equivalent to the
integrable boundary SSG theory. An analysis of finding the general
form of integrable boundary potential will be reported elsewhere.
This can be done by use of the boson-fermion duality
relations (\ref{bf-relation1}) and (\ref{bf-relation2}). Experience suggests
us to make the following ansatz for the general form of the boundary
GN model lagrangian, 
\beq
{\cal L}_{bGN}=-\frac{1}{2}\sum_{i=1}^3\bar{\chi}_ii\g^\mu\p_\mu\chi_i
   -g\lt(\sum_{i=1}^3\bar{\chi}_i\chi_i\rt)^2
   + i\d(x)\lt(\frac{g'}{4}\bar{\chi}_3\chi_3+g''(\bar{\chi}_1\chi_1
   +\bar{\chi}_2\chi_2)\rt ).\label{b-gn-gen}
\eeq

We determine the constants 
$g'$ and $g''$ by demanding that the boundary potential be integrable. Following
the same prescription as in \cite{Wit78}, one obtains from (\ref{b-gn-gen})
\bea
\cL_{bGN}&=&-\frac{1}{2}(\p_\mu\varphi)^2-\frac{1}{2}\bar{\chi}_3i\g^\mu\p_\mu
    \chi_3-2A^2:\cos\rho\varphi:^2-iA\rho\,
    \bar{\chi}_3\chi_3:\cos\rho\varphi:\no\\
& &+\d(x)\lt (i\frac{g'}{4}\bar{\chi}_3\chi_3+g'':\cos\rho\varphi:\rt),
\eea
where $\varphi,~\rho$ and $A$ are given by (\ref{funny-constants}).

Comparing with the integrable
boundary SSG theories, one sees that in order for the boundary  GN
to be integrable the constants $g'$ and $g''$ should take either of
the two values:
\bea
&& 1)~~~~g'=M,~~g''=\L;\no\\
&& 2)~~~~g'=\pm 1,~~g''=\pm\frac{2A}{B}=\pm\frac{g(1-4g/\pi)}{\pi-8g^2/\pi}.
   \label{g'-g''}
\eea
Thus one has two alternative classes of integrable boundary  GN model which are
defined by
\bea
1)~~~\cL_{bGN}&=&-\frac{1}{2}\sum_{i=1}^3\bar{\chi}_ii\g^\mu\p_\mu\chi_i
   -g\lt(\sum_{i=1}^3\bar{\chi}_i\chi_i\rt)^2\no\\
& &   + i\d(x)\lt(\frac{M}{4}\bar{\chi}_3\chi_3+\L (\bar{\chi}_1\chi_1
   +\bar{\chi}_2\chi_2)\rt ), ~~~M(\neq\pm 1),
   ~\L~{\rm arbitary};\label{b-gn-non-susy}\\
2)~~~\cL_{bGN}&=&-\frac{1}{2}\sum_{i=1}^3\bar{\chi}_ii\g^\mu\p_\mu\chi_i
   -g\lt(\sum_{i=1}^3\bar{\chi}_i\chi_i\rt)^2\no\\
& &\pm i\d(x)\lt (\frac{1}{4}\bar{\chi}_3\chi_3+
   \frac{g(1-4g/\pi)}{\pi-8g^2/\pi}\,(\bar{\chi}_1\chi_1
   +\bar{\chi}_2\chi_2)\rt).\label{b-gn-susy}
\eea

For the special case $M=4\L$ the model 1) is reduced to 
\beq
1')~~~\cL_{bGN}=-\frac{1}{2}\sum_{i=1}^3\bar{\chi}_ii\g^\mu\p_\mu\chi_i
   -g\lt(\sum_{i=1}^3\bar{\chi}_i\chi_i\rt)^2
   + i\d(x)\L\sum_{i=1}^3\bar{\chi}_i\chi_i
\eeq
This model preserves the $O(3)$ symmetry of the bulk theory. On the other
hand, the $O(3)$ symmetry is always broken in the model 2).

\sect{Discussions}

We have derived the general form of boundary interactions $\cB(\psi)$ by finding
all solutions of (\ref{t4-t4bar}) in the MT model. We are then able to
discuss the boson-fermion duality between the boundary MT model and the
boundary SG model. It is very curious that the integrable boundary
potential for the MT model contains a larger number of free parameters than
that for the SG model.

To study the duality between the few classes of boundary SSG model and those
of the $O(3)$ GN model in details, we have to go beyond the present
investigation in section \ref{gn-integrability} and  to
carry out an analysis similar to the one in section
\ref{mt-integrability}. It is of interest to do the same for the
$SU(2)$-invariant Thirring model. We wish to return to these problems in the
future.

The SG/MT model is obtained by taking the continuum limit of 
the spin-$\frac{1}{2}$ XYZ chain with appropriate choice of the coupling
constants in the latter model \cite{Lut76}. 
It is very interesting to see whether
the boundary SG/MT model can be obtained from the boundary XYZ spin chain
\cite{Hou93} by taking the continuum limit. 
This problem has been discussed in \cite{Fen94} and is under investigation.

\vskip.3in
The authors are grateful to Ed Corrigan
for a careful reading of the manuscript. T.I. wishes to thank Choonkyu
Lee for a discussion on the boson-fermion duality while he was visiting Seoul
supported by the Japan-Korea exchange program of JSPS. 
This work was partially supported by the Grant-in-Aid
for Scientific Research, priority area 231,
from the Ministry of Education, Science and Culture of Japan.
H.K. is supported by the Yukawa Memorial Foundation.
Y.Z.Z. is supported by the Japan Society for the
Promotion of Science (JSPS). 

When writing this paper, we saw a recent preprint \cite{Gao95} in which
a boundary MT model corresponding to the special case of one of our solutions,
(\ref{solution3}), is derived via fermionization of the boundary SG
theory by using the same method as in \cite{LeC95}.

%\newpage


\vskip.3in
\begin{thebibliography}{99}
\bibitem{Mon77} C. Montonen, D. Olive, {\em Phys.Lett.} {\bf B72} (1977) 117;\\
   P. Goddard, J. Nuyts, D. Olive, {\em Nucl.Phys.} {\bf B125} (1977) 1;\\
   N. Seiberg, E. Witten, {\em Nucl.Phys.} {\bf B426} (1994) 19. 
\bibitem{Col75}  S. Coleman, {\em Phys.Rev.} {\bf D11} (1975) 2088;\\ 
   S. Mandelstam, {\em Phys.Rev.} {\bf D11} (1975) 3026.
\bibitem{Wit84} E. Witten, {\em Commun.Math.Phys.} {\bf 92} (1984) 455;\\
   M.B. Halpern, {\em Phys.Rev.} {\bf D12} (1975) 1684, and {\bf D13} 
   (1976) 337.
\bibitem{Che84} I. Cherednik,  {\em Theor.Math.Phys.} {\bf 61} (1984) 35.
\bibitem{Car89}  J. Cardy, {\em Nucl.Phys.} {\bf B324} (1989) 581.
\bibitem{Gho94}   S. Ghoshal, A.B. Zamolodchikov, {\em Int.J.Mod.Phys.}
   {\bf A9} (1994) 3841.
\bibitem{Cor94}  E. Corrigan, P.E. Dorey, R.H. Rietdijk, R. Sasaki,
   {\em Phys.Lett.} {\bf B333} (1994) 83.
\bibitem{Sal95}  H. Saleur, S. Skorik, {\em J.Phys.} {\bf A:} {\em Math.Gen.}
   {\bf 28} (1995) 6605.
\bibitem{Ina95} T. Inami, S. Odake, Y.-Z. Zhang,  {\em Phys.Lett.} {\bf B359}
   (1995) 118.
\bibitem{Ahn95}   C. Ahn, W.-M. Koo, preprint hep-th/9509056.
\bibitem{Zam79}  A.B. Zamolodchikov, Al.B. Zamolodchikov, {\em Ann.Phys.}
   {\bf 120} (1979) 253.
\bibitem{Wit78}  E. Witten,  {\em Nucl.Phys.} {\bf B142} (1978) 285;\\
    H. Aratyn, P.H. Damgaard, {\em Nucl.Phys.} {\bf B241} (1984) 253.
\bibitem{Bur94}  C.P. Burgess, F. Quevedo, {\em Nucl.Phys.} {\bf B421} (1994)
   373;\\ P.H. Damgaard, H.B. Nielsen, R. Sollacher, {\em Phys.Lett.}
   {\bf B296} (1992) 132.
\bibitem{Ber76}  B. Berg, M. Karowski, H.J. Thun, {\em Phys.Lett.}
   {\bf 64B} (1976) 286;\\ P.P. Kulish, E.R. Nissinov, {\em JETP Lett.}
   {\bf 24} (1976) 247;\\
   R. Flume, P.K. Mitter, N. Papanicolaou, {\em Phys.Lett.} {\bf 64B}
   (1976) 289.
\bibitem{Bow95} P. Bowcock, E. Corrigan, P.E. Dorey, R.H. Rietdijk,
   {\em Nucl.Phys.} {\bf B445} (1995) 469.
\bibitem{LeC95} M. Ameduri, R. Konik, A. LeClair, {\em Phys.Lett.} {\bf B354}
   (1995) 376.
\bibitem{Lut76} A. Luther, {\em Phys.Rev.} {\bf B14} (1976) 2153.
\bibitem{Hou93} B.-Y. Hou, R.-H. Yue, {\em Phys.Lett.} {\bf A183} (1993) 169;\\
   T. Inami, H. Konno, {\em J.Phys.} {\bf A:} {\em Math.Gen.}
   {\bf 27} (1994) L913.
\bibitem{Fen94} P. Fendley, H. Saleur, {\em Nucl.Phys.} {\bf B428}[FS] (1994)
    681;\\  M.T. Grisaru, L. Mezincescu, R.I. Nepomechie, {\em J.Phys.}
    {\bf A:} {\em Math.Gen.} {\bf 28} (1995) 1027.
\bibitem{Gao95} H.-B. Gao, Z.-M. Sheng,  preprint hep-th/9512011.


\end{thebibliography}
\end{document}